\documentclass[twocolumn]{aastex631}
\usepackage{amsmath}	% Advanced maths commands
\usepackage{amssymb}	% Extra maths symbols
\usepackage{hyperref}
\usepackage{newtxtext,newtxmath}
\usepackage{float}
\usepackage{longtable}
\usepackage{tabularx}
\usepackage{float}
\usepackage{comment}
\usepackage{ragged2e}
\usepackage{booktabs}

\usepackage{todonotes}
\usepackage{multirow}
\usepackage{threeparttable}

\shortauthors{Bora et al.}
\begin{document}

\title{Multi-wavelength polarization degree dependence of Mrk 421 for different particle energy distributions}

\author[0000-0002-1520-057X]{Hritwik Bora}
\affiliation{Department of Physics, Tezpur University, Tezpur-784028, India}

\author[0000-0002-7609-2779]{Ranjeev Misra}
\affiliation{Inter-University Centre for Astronomy and Astrophysics, Post Bag 4, Ganeshkhind, Pune - 411007, India}

\author[0000-0003-1715-0200]{Rupjyoti Gogoi}
\affiliation{Department of Physics, Tezpur University, Tezpur-784028, India}

%\author{August Muench}
%\affiliation{American Astronomical Society \\
%1667 K Street NW, Suite 800 \\

%\collaboration{20}{(AAS Journals Data Editors)}

%\author{F.X Timmes}
%\affiliation{Arizona State University}
%\affiliation{AAS Journals Associate Editor-in-Chief}

%\altaffiliation{AASTeX v6+ programmer}

%\affiliation{AAS Director of Publishing}
%\affiliation{American Astronomical Society \\
%1667 K Street NW, Suite 800 \\



\begin{abstract}

We report on the constraints obtained from the multi-wavelength polarization degree and energy spectra measurements for the High Synchrotron Peaked Blazar (HBL) source \object{Mrk 421} using different forms of the electron energy distribution. We find that the optical and X-ray data are consistent with the electron energy distribution being a broken power-law (BPL) or other distributions, with all electrons experiencing the same magnetic field configuration. This is in contrast to other HBL sources like the Mrk 501, where, for the BPL distribution, the X-ray producing higher energy electrons need to exist in a more ordered magnetic field environment than the optical producing lower energy ones. This is because the break energy of the spectrum produced by the BPL distribution for \object{Mrk 421} lies below the IXPE energy band. We discuss the implications of the findings. 

\end{abstract}


\section{Introduction} \label{sec:intro}

Blazars are the subclasses of Active Galactic Nuclei (AGN) with relativistic jets aligned close to our line of sight, leading to Doppler boosted non-thermal emission across the electromagnetic spectrum. Their broadband spectral energy distributions (SED) exhibit two broad humps. The first hump is synchrotron emission peaking at optical/X-ray regime, and the second is usually attributed to inverse Compton processes (either synchrotron self-Compton or external Compton) or to hadronic particle interactions extending to $\gamma$-ray regime \citep{urry1998multiwavelength,massaro2004log}.

The launch of the Imaging X-ray Polarimetry Explorer (IXPE) in December 2021 has opened a new window for blazar studies \citep{Zhang_2021, 2022Natur.611..677L, 2023NatAs...7.1245D, Hu_2024}. In particular, it allows direct measurement of the polarization of X-ray synchrotron emission from HBLs. 

Polarization adds a critical probe of the jet’s magnetic field and particle dynamics. Recent models specifically address the energy-stratified shock scenario \citep{2022Natur.611..677L, 2023NatAs...7.1245D, 2024A&A...681A..12K}. If a shock accelerates electrons and a helical/toroidal field is present, the highest-energy electrons radiate X-rays near the shock front before cooling. In that compact zone the field is relatively ordered (shock-compressed), leading to high degree of polarization ($\Pi$). In addition, if the diffusion time scale is comparable to the cooling one, then the higher energy particles will exist preferentially near the shock and hence experience a more ordered magnetic field. This is in contrast to the standard assumption made during broadband spectral fitting that the emission region is homogeneous and the particles of all energies experience the same magnetic field \citep{10.1093/mnras/stae706}. It has been shown that intrinsically curved particle energy distributions, such as log-parabola or energy-dependent acceleration models, can naturally reproduce higher X-ray polarization relative to optical bands even in a uniform magnetic field configuration \citep{Misra_2025}. This introduces a fundamental degeneracy in interpreting polarization measurements, as both magnetic field structure and particle distribution shape can influence the observed polarization properties. This energy-dependent polarization provides crucial insights into the structure and ordering of magnetic fields as well as the nature of particle acceleration processes in the jet.
However, this interpretation implicitly assumes that the underlying particle energy distribution follows a simple power-law or broken power-law form \citep{Anderhub_2009, Abe_2023, 2024A&A...685A.117M}. Broadband SED modeling of blazars, including Mrk 421, has shown that intrinsically curved particle energy distributions, such as log-parabola forms, can also successfully reproduce observed spectra \citep{2004A&A...413..489M, 2007A&A...466..521T}. 

Mrk 421 ($z \sim 0.031$) is a nearby HBL source detected in TeV, making it a prime target for multi-wavelength spectro-polarimetric studies. \cite{2023NatAs...7.1245D} reports the first X-ray polarization detection of Mrk 421. Observing a moderate state in May 2022, they measured a 2–8 keV polarization degree $\Pi_x$ = $15 \pm 2\%$ with EVPA $\Psi_x$ = $35^\circ \pm 4^\circ$. Simultaneous multi-band polarization showed much lower $\Pi$ in infrared/optical/mm ($\leq$ 5\%) \citep{2023NatAs...7.1245D, 2024A&A...681A..12K, Bharathan_2024}. There is also radio polarization measurement for this source \citep{2023NatAs...7.1245D}. 

Thus, Mrk 421 is another source where one can study multi-wavelength polarization and energy spectral data to infer which particle distributions are consistent with the data and, in particular, to quantify whether different-energy electrons experience different magnetic field configurations, i.e., the energy-stratified shock scenario. Moreover, the radio observations may provide further clues to the nature of the source.

In this work, we investigate the polarization properties of Mrk 421 by modeling synchrotron emission from particles with different energy distributions in the presence of both ordered and turbulent magnetic field components. By comparing theoretical predictions with multi-wavelength spectral and polarization observations, we aim to assess whether a unified magnetic field configuration can explain the data or whether an electron energy-dependent magnetic structure is required. This approach provides a framework for better understanding the interplay among particle acceleration, magnetic field topology, and radiation processes in relativistic jets.

The paper is organized as follows: Sections \ref{sec:obs} and \ref{sec:sec3} detail the Data analysis and formulism to estimate the degree of polarization for arbitrary particle distribution. Section \ref{sec:sec4} and \ref{sec:pd} discuss the modeling and the electron particle distribution used. Section \ref{sec:results} presents the results, and Section \ref{sec:Discussion} summarizes and discusses the findings. 

\section{Observations and Data Analysis}
\label{sec:obs}

\subsection{X-ray Polarization}

The Imaging X-ray Polarimetry Explorer (IXPE), launched by NASA in December 2021, is the first mission dedicated to X-ray polarimetry. It comprises three identical grazing-incidence telescopes operating in the $2$--$8$ keV band. Equipped with Gas Pixel Detectors, IXPE enables sensitive polarization measurements, proving highly effective in probing magnetic field geometry and emission mechanisms \citep{2022JATIS...8b6002W}.

IXPE observed Mrk 421 several times. For our work, we used the first polarization observation conducted during May 2022 (reported in Table \ref{tab:obs}). \cite{2023NatAs...7.1245D} reports a significant detection of the degree of polarization $\sim$ 15\% and the angle of polarization of $\sim 35^\circ$. In two subsequent pointing during June 2022, the polarization remained undetected (through time-averaged methods). The observations of Mrk 421 were complemented with different observatories mentioned below.

\subsection{Optical Polarization}

For the optical polarization, we have adopted the PD of $\sim 2.9 \pm 0.1 \%$ (from Nordic Optical Telescope) and $\sim 3.1 \pm 0.2 \%$ (from KANATA) are in the optical band. Note that the polarization points from NOT are corrected, while those from KANATA are not corrected for the unpolarized light from the host galaxy \citep{2023NatAs...7.1245D}. We loaded the polarization data points detected in different wavelengths by converting them into \texttt{XSPEC} (version 12.11.1) \citep{1996ASPC..101...17A} readable PHA format with the help of the tool. \texttt{ftflx2xsp}.      

\subsection{NuSTAR}
The Nuclear Spectroscopic Telescope Array (NuSTAR) is the first focusing hard X-ray observatory in orbit, operating in the 3--79 keV energy range.
Launched by NASA on June 13, 2012, NuSTAR significantly extended imaging and spectroscopic capabilities beyond $\sim$10 keV. The instrument comprises two focal plane modules, FPMA and FPMB \citep{Harrison_2013}. The NuSTAR observations of Mrk~421 analyzed in this study are listed in Table \ref{tab:obs}.

Data reduction was carried out using \texttt{NUPIPELINE} within the \texttt{HEASOFT-6.28} to produce cleaned event files. The \texttt{XSELECT~V2.4k} package was employed to visualize the cleaned event files (\textit{cl.evt}) in \texttt{ds9}. Source spectra was extracted from circular regions of $30\arcsec$ while background spectra was obtained from a source-free region of radius 60\arcsec. Spectrum files (PHA), ancillary response files (ARF), and response matrix files (RMF) were generated using the \texttt{NUPRODUCTS} task. The resulting spectra were grouped with \texttt{GRPPHA} to a minimum of 30 counts per bin, ensuring statistically reliable spectral fitting.

\subsection{\textit{Swift}-XRT}
We analyzed \textit{Swift}-XRT observations that are quasi-simultaneous with the NuSTAR data. The XRT instrument operates in the 0.2--10 keV energy band with an effective area of approximately 135 cm$^2$. All observations from both NuSTAR and \textit{Swift}-XRT were retrieved from NASA's HEASARC archive\footnote{\url{https://heasarc.gsfc.nasa.gov/}}. Cleaned event files were produced using \texttt{XRTPIPELINE} under \texttt{HEASOFT}, and \texttt{XSELECT} was used to inspect the event files (\textit{cl.evt}) in \texttt{ds9}.

All observations were acquired in Photon Counting (PC) mode. Source and background spectra were extracted from circular regions of 10 and 20 pixel radius, respectively. ARF and RMF files were generated using the \texttt{xrtmkarf} and \texttt{quzcif} tools. As with NuSTAR, the \textit{Swift}-XRT spectra were grouped to a minimum of 30 counts per bin prior to spectral analysis.

\subsection{\textit{Swift}-UVOT}
Simultaneous ultraviolet and optical observations were obtained using the \textit{Swift}-UVOT instrument. The instrument covers both optical and UV bands through three optical filters (B, V, U) and three UV filters (UVW1, UVW2, UVM2) \citep{2005SSRv..120...95R}; observations used here were available in the UVW1, UVW2, and UVM2 filters. Source and background fluxes were extracted from apertures of 7 and 14 pixel radius, respectively.

Magnitudes were computed in the AB system using the \texttt{UVOTSOURCE} tool and subsequently corrected for Galactic extinction via the standard relation $A_{V}/E(B-V) = 3.1$, where $A_{V}$ denotes the Galactic extinction and $E(B-V)$ the color excess \citep{schlafly2011measuring}. For Mrk 421, we adopted $E(B-V) = 0.013$~mag\footnote{\url{https://irsa.ipac.caltech.edu/applications/DUST/}}. Flux densities were derived from the corrected magnitudes using photometric zero points and conversion factors from \citet{2011AIPC.1358..373B} and \citet{10.1093/mnras/stw1516}. The resulting flux values and corresponding photon energies were then converted into a PHA format compatible with \texttt{XSPEC} using the \texttt{ftflx2xsp} tool.

\begin{table}
\renewcommand{\arraystretch}{1.5}
\centering
\caption {Summary of NuSTAR, Swift-XRT/UVOT, and IXPE Observations of Mrk 421 for MJD = 59703.}
\label{tab:obs}
\begin{tabular}{lccc}
\hline
 & \textbf{NuSTAR} & \textbf{Swift-XRT/UVOT} & \textbf{IXPE} \\
\hline
Obs ID & 60701031002 & 00031540032 & 01003701 \\

\multirow{2}{*}{Date and Time} 
& 2022-05-04 & 2022-05-03 & 2022-05-04 \\
& T14:26:09  & T20:12:41  & T10:00:28 \\

Exposure (ks) 
& 24.73  & 0.9 / 0.9 & 96.5 \\
\hline
\end{tabular}
\end{table}

\section{Polarization from arbitrary particle distributions}\label{sec:sec3}

The expected polarization of a synchrotron emitting an arbitrary particle distribution is given in detail in \cite{Misra_2025}. In this section, we provide a brief version of the description for ease of reading and understanding.

The single particle synchrotron photon spectrum from an electron with Lorentz factor $\gamma$ in a random magnetic field $B_R$ is \citep[e.g.][]{2009herb.book.....D},

\begin{equation*}
N_R (x_R) = \frac{2 \pi e^2}{\sqrt{3} \gamma^2 c h} x_R G\!\left(\frac{x_R}{2}\right)
\label{spec_rand}
\end{equation*}

where,
\begin{equation*}
    G(x) = K_{4/3}(x)K_{1/3}(x)-\frac{3}{5}x[K_{4/3}^2 (x)-K_{4/3}^2(x)]
\end{equation*} 

\noindent and $\nu_R = \frac{3e}{4\pi m_e c}\gamma^2 B_R$. For an ordered field $B_O$, the pitch-angle-integrated  spectrum is \citep[e.g.][]{1986rpa..book.....R},

\begin{equation}
  N_O(x_O) = \frac{2 \pi e^2 }{\sqrt{3} \gamma^2 c h} \int_{x_O}^{\infty} K_{5/3} (x_{O}') dx_O'
\label{spec_order}
\end{equation}

\noindent where, $x_O = h\nu/h\nu_O$ and $\nu_O = \frac{3e}{4\pi m_e c}\gamma^2 B_O\sin\theta$. One can verify 
that $P_R = (2/3)P_O$ when $B_R = B_O$ \citep{1986rpa..book.....R}.

We have adopted the spectro-polarimeteric relations from \cite{Misra_2025}. In order to use it is convenient to introduce the variable 
$\xi = \sqrt{\mathbb{C}}\,\gamma$, where $\mathbb{C} = \frac{3eh}{4\pi m_e c}\frac{\delta_D B_R}{(1+z)}$ \citep{Hota_2021, Khatoon_2022, 10.1093/mnras/stae706, hota2024multiwavelengthstudyextremehighenergy, tantry2024probingbroadbandspectralenergy, Bora_2026}, so that the predicted spectrum depends only on the particle distribution $n(\xi)$ and not on the degenerate parameters $\delta_D$ and $B_R$ individually. For a system with both ordered and random fields, the observed spectrum is, 

\begin{equation}
S_T(E) = S_R(E) + S_O(E)    
\label{spec1}
\end{equation}

Where, 

\begin{equation*}
    S_R(E) = \frac{\delta_D^3(1+z)V}{d_L^2\sqrt{\mathbb{C}}} \int_{\xi_{min}}^{\xi_{max}} (1-f_O) N_R (\frac{E}{\xi^2}) n(\xi) d\xi
\end{equation*}
and 
\begin{equation*}
    S_O(E) = \frac{\delta_D^3(1+z)V}{d_L^2\sqrt{\mathbb{C}}} \int_{\xi_{min}}^{\xi_{max}} f_O N_O (\frac{E}{\eta \xi^2}) n(\xi) d\xi
\end{equation*}

\noindent The polarization degree $\Pi_O$ from the ordered field is \citep{1986rpa..book.....R} 

\begin{equation*}
    \Pi_O(E) = \frac{\int_{\xi_\text{min}}^{\xi_\text{max}} K_{2/3} (\frac{E}{\eta \xi^2})f_o n(\xi) d\xi}{\int_{\xi_\text{min}}^{\xi_\text{max}} \left[ \int_{E/\eta \xi^2}^{\infty} K_{5/3} (x') dx' \right] f_o n(\xi) d\xi}
\end{equation*}

with $\eta = \frac{B_O\sin\theta}{B_R}$ and $f_O$ the volume fraction associated with the ordered field.

The observed polarization degree is given by \citep{Misra_2025},

\begin{equation}
    \Pi_T(E) = \frac{S_O(E)}{S_T{(E)}} \Pi_O(E)
    \label{pol}
\end{equation}

\section{Spectro-polarimetric Modeling} \label{sec:sec4}

The multi-wavelength observation on 4th May 2022 of Mrk 421 reveals the degree of X-ray polarization ($15 \pm 2\%$) was significantly higher than the median intrinsic optical and radio polarization degree \citep{2023NatAs...7.1245D}. We do not consider the high-energy \textit{Fermi}-LAT data since the focus here is on the synchrotron emission. For different particle energy distributions, we formally fit the optical and X-ray spectra, and the polarization values using the formulation described in Section \ref{sec:sec3}, with emphasis on the polarization degree predictions. The polarization degree as a function of energy is converted to an \texttt{XSPEC} \citep{1996ASPC..101...17A} readable format and then both the spectral energy distribution and polarization degree are fitted within the \texttt{XSPEC} environment. A local \texttt{XSPEC} convolution model \texttt{synpol} was created such that if a flag is set to zero, then the output is the broadband synchrotron spectrum (using Equation \ref{spec1}), and if set to one it outputs the polarization degree (using Equation \ref{pol}). The model is then convolved with different particle distributions to obtain a fit with parameter values and errors. As mentioned below, the spectral energy distribution is rather insensitive to the parameters representing the ordered magnetic field, namely $f_O$ and $\eta$. For the spectral fit, we have used the \texttt{Tbabs} model \citep{Wilms_2000} to account for the Galactic absorption. The hydrogen column density for the X-ray observations, ${N_\text{H}} = 1.34 \times 10^{20} $ $\text{cm}^{-2}$ was considered and kept fixed which was obtained in the LAB survey \citep{2005A&A...440..775K}. For the UV observations, the $N_\text{H}$ value was fixed at 0 as the fluxes were de-redenned initially. The errors are estimated at a 90\% confidence level with the standard \texttt{XSPEC} method of $\chi^2$error calculation\footnote{\url{https://heasarc.gsfc.nasa.gov/xanadu/xspec/manual/XSerror.html}}. In \texttt{XSPEC}, the model has been defined as,

$$constant \otimes tbabs \otimes synpol \otimes n(\xi)$$

\section{Particle Distributions}\label{sec:pd}
\subsection{{Logparabola model}}
The particle density for a log-parabolic model is given by,

\begin{equation}\label{lp} 
    n({\xi})=K \left (\frac{\xi}{\xi_r} \right)^{- \alpha - \beta \text{log} \left(\frac{\xi}{\xi_r} \right)} 
\end{equation}

\noindent In this expression, $\alpha$ represents the particle spectral index at the reference energy $\xi_r$, while $\beta$ and $K$ denote the spectral curvature parameter and normalization, respectively. During the spectral fit, $\xi_r^2$ was fixed at 1 keV, while $\alpha$, $\beta$, and the normalization $K$ were treated as free parameters.

\subsection{Broken Power Law}
We applied a Broken Power Law (BPL) distribution to model the particle spectrum,

\begin{equation}\label{bpl}
n(\xi)=
\begin{cases}
    K (\xi/1\sqrt{{\text{keV}}})^{-p} & \text{for} \hspace{3pt} \xi < \xi_\text{break} \\
    K \xi^{q-p}_\text{break}(\xi/1\sqrt{{\text{keV}}})^{-q} & \text{for} \hspace{3pt} \xi > \xi_\text{break}  
\end{cases}
\end{equation}

\noindent Here, $\xi_\text{break}$ represents the break energy, $p$ is the electron spectral index for $\xi < \xi_\text{break}$, and $q$ is the electron spectral index for $\xi > \xi_\text{break}$. The transformation is defined by $\xi = \gamma \sqrt{\mathbb{C}}$. This broken power-law particle distribution is considered valid only within the range $\xi_\text{min} < \xi < \xi_\text{max}$, where $\xi_\text{min} = \gamma_\text{min} \sqrt{\mathbb{C}}$ and $\xi_\text{max} = \gamma_\text{max} \sqrt{\mathbb{C}}$.

\subsection{Energy Dependent Diffusion Model (EDD)}

In this model, diffusion occurs in a region with a tangled magnetic field, causing the escape timescale to depend on the electron's gyration radius. This makes the escape timescale ($\tau_\text{esc}$) energy-dependent, given by

\begin{equation} \tau_\text{esc} = \tau_{\text{esc},R} \left(\frac{\gamma}{\gamma_R} \right)^{-k} \end{equation}

\noindent where $\tau_{\text{esc},R}$ is the escape timescale when the electron energy is $\gamma_{R}mc^2$, and $k$ is the index describing the power-law dependence on energy.

The escape timescale $\tau_\text{esc}$ cannot exceed the free-streaming limit, so this relationship is valid only for $\gamma < \gamma_R$, where $\gamma_R$ is the energy at which this limit is reached. In steady state the particular distribution is determined by the balance of the escape rate with the acceleration process characterized by (a non-energy dependent) time-scale $\tau_{acc}$.

Assuming $\gamma_R$ is much larger than any $\gamma$ of interest and neglecting synchrotron losses, the resulting electron energy distribution is,

\begin{equation} n(\xi) = Q_0 \tau_\text{acc} \sqrt{\mathbb{C}} \xi^{-1} \exp \left[ -\frac{n_R}{k} \left( \left(\frac{\xi}{\xi_R} \right)^k - \left( \frac{\xi_0}{\xi_R} \right)^k \right) \right] \end{equation}

\noindent where $\xi_R = \sqrt{\mathbb{C}} \gamma_R$, $\xi_0 = \sqrt{\mathbb{C}} \gamma_0$, and $\eta \equiv \frac{\tau_\text{acc}}{\tau_{\text{esc},R}}$. Where $\gamma_0$ is the minimum injection energy of the electron.

The distribution can be conveniently rewritten as,

\begin{equation} n(\xi) = K \xi^{-1} \exp \left[-\frac{\psi}{k} \xi^{k} \right] \end{equation}

\noindent where $K$, $\psi$, and $k$ are free parameters. It can be shown that,

$$\psi = \eta_R (\mathbb{C}\gamma^{2}_{R})^{-k/2}=\eta_R \xi_{R}^{-k}$$

and that the normalization is given by,
    
$$K = Q_0 \tau_{acc} \text {exp} \left [\frac{n_R}{k} \left(\frac{\xi_0}{\xi_{R}} \right)^k \right]$$

\begin{table*}
\caption {\label{tab:bestfit} The best-fit parameters obtained for all the particle distributions considered. The errors have been calculated at 90\% confidence level.}
\renewcommand{\arraystretch}{1.55} % Increase row height by 50%
\normalsize
%\toprule
%\toprule
\centering
%\resizebox{\textwidth}{!}{
%\begin{tabularx}{\textwidth}{ccccccccc}
\begin{tabular*}{\textwidth}{@{\extracolsep{\fill}}lcccccc}
\hline
\multicolumn{7}{c}{\textbf{Log Parabola}} \\  
%\cmidrule{2-7}
\hline
 & $\alpha$  &  $\beta$  &  $\eta$  &   $f_{o}$   & $\chi^{2}_\text{red}(\text {dof})$ \\
%\cmidrule{2-7}
 & 4.39$^{+0.05}_{-0.05}$ &  1.51$^{+0.09}_{-0.08}$ &  2.0$^{+0.3}_{-0.3}$ & 0.034$^{+0.007}_{-0.005}$ & 1.12(592)\\ 
%\cmidrule{2-7}
\hline
\multicolumn{7}{c}{\textbf{Broken Power Law}} \\  
%\cmidrule{2-7}
\hline
 & $p$ & $q$ & $\xi_\text{brk}$ ($\sqrt{\text{keV}})$ & $\eta$ & $f_{o}$ & $\chi^{2}_\text{red}(\text {dof})$ \\
 & $2.47 ^{+0.09}_{-0.12}$  & $5.37^{+0.06}_{-0.05}$ & $0.77^{+0.04}_{-0.04}$ & $4.4^{+0.5}_{-0.4}$ & $0.004^{+0.002}_{-0.001}$ & 1.18 (591)\\
\hline
\multicolumn{7}{c}{\textbf{EDD}} \\  
%\cmidrule{2-7}
\hline
& $\psi$ & $k$ & $\eta$ & $f_{o}$ & $\chi^{2}_\text{red}(\text {dof})$ \\
%\cmidrule{2-7}
& 3.24$^{+0.06}_{-0.05}$ & 0.45$^{+0.03}_{-0.02}$ &  3.0$^{+0.5}_{-0.5}$ & 0.013$^{+0.004}_{-0.003}$ & 1.14(592)\\ 
\hline
\end{tabular*}

\end{table*}

\begin{figure*}[ht!]
\centering
    \includegraphics[width=.5\textwidth]{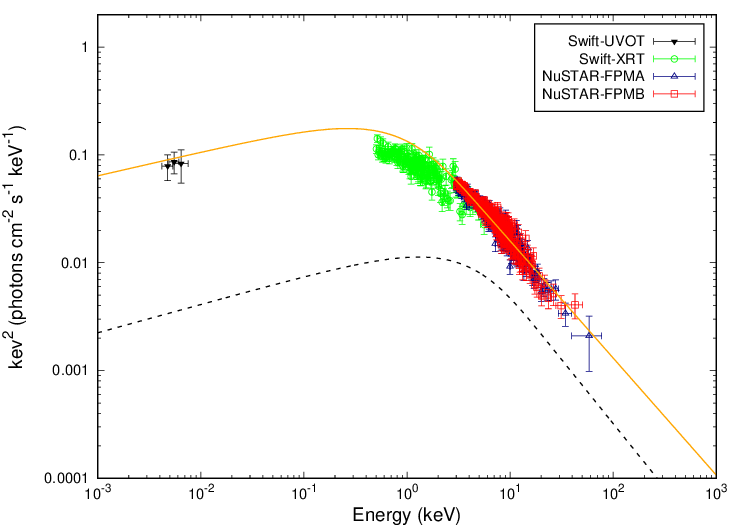}\hfill
    \includegraphics[width=.5\textwidth]{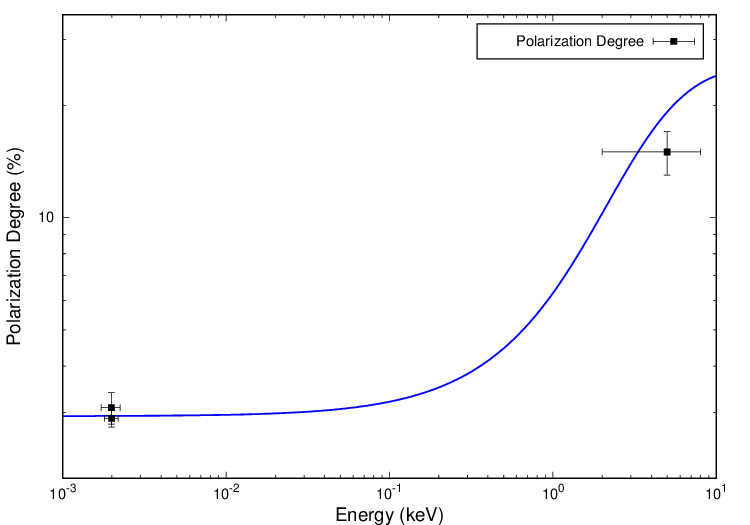}\hfill
    \includegraphics[width=.5\textwidth]{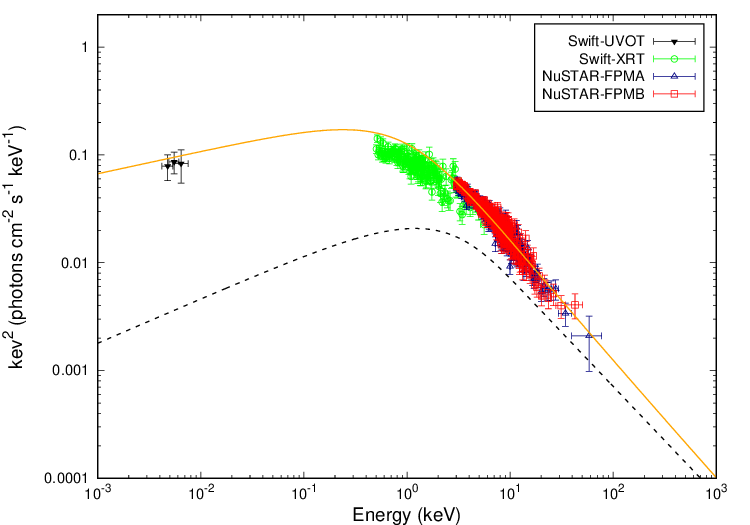}\hfill
    \includegraphics[width=.5\textwidth]{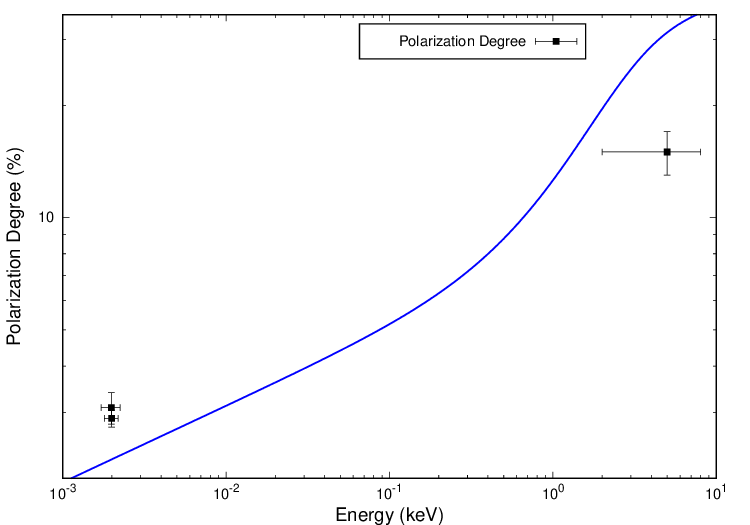}\hfill
    \caption{The left panels present broadband SED, while the right panels show the polarization degree in the optical and X-ray bands of Mrk 421 \citep{2023NatAs...7.1245D}. The solid lines represent the model prediction from a Broken-Powerlaw (BPL) particle energy distribution. Note that the model energy spectra are  unabsorbed, whereas the X-ray data points are  with absorption. The dotted lines represent synchrotron emission from the ordered magnetic field components. The top panel represents model spectrum and polarization for the case when all particles experience the same magnetic field configurarion. The bottom panel represents the case when the higher energy particles exist in a more ordered magnetic field as required for fitting Mrk 501 data.}
    \label{fig1}
\end{figure*}

\section{Results}\label{sec:results}

We performed a multi-wavelength spectro-polarimetric study of Mrk 421 in the optical, UV, and X-ray bands using instruments NOT, KANATA, \textit{Swift}-UVOT/XRT, NuSTAR, and IXPE. The observations were fitted with a local convolution model in \texttt{XSPEC} for Logparabola, Broken Power Law, and Energy Dependent Diffusion particle distribution. We have found that all the models fit the multi-wavelength spectro-polarimetric observations from optical to X-ray bands. Table \ref{tab:bestfit} reports best-fit parameters for polarization data spanning from optical to X-ray energies for LP and EDD models, whereas for the BPL distribution, the fitting includes polarization data from radio to X-ray energies. In the left panel of Figure \ref{fig1}, the solid line represents the best-fit spectrum given by $S_T(E)$, which shows the expected polarization for the parameters $f_o$ and $\eta$, that bring the model closest to the observed ones. However, note that in all the figures, we have shown an unabsorbed model, hence there is a deviation of the model from the data at lower energies. The dotted line represents the emission when the magnetic field is ordered, given by $S_O (E)$ for an ordered field volume fraction $f_o$ and ratio $\eta = \frac{B_O \sin\theta}{B_R}$. The right panel of the Figure \ref{fig1} shows the polarization degrees in the optical and X-rays bands as reported by \cite{2023NatAs...7.1245D}, with the solid line being the predicted energy polarization degree $\Pi_T(E)$. 

\subsection{Optical and X-ray Polarization}

As shown in the top panel of Figure \ref{fig1}, for the BPL distribution, we could fit both the optical and X-ray polarizations under the assumption that the emitting particles exist within the same magnetic field configuration. This behavior may arise because, in the present case, the expected polarization degree increases toward higher energies beyond the break energy. This is in contrast to the findings of \cite{Misra_2025}, where the break energy was reported to lie beyond the polarization detection range, preventing such an increase from being observationally constrained.

To further investigate the role of the magnetic field structure, we explored whether the observed polarization data could still be reproduced by varying the degree of magnetic field ordering. To do this, for simplicity, we could choose this dependence to be a power law such that $f_o(\xi) = f_{ox} \xi^\chi$. We first adopted a fixed value of $\chi = 0.21$ from \cite{Misra_2025} and attempted to model the polarization measurements. However, as illustrated in the bottom-right panel of Figure \ref{fig1}, the model fails to adequately describe the data. In particular, the predicted polarization degree remains significantly below the observed optical polarization and exhibits a sharp increase near the break energy, resulting in a poor overall agreement with the observations.

Subsequently, we allowed $\chi$ to vary as a free parameter in the fitting procedure. We found that a good fit requires $\chi \lesssim 0.02$, indicating a tight constraint on the electron energy dependence of the magnetic field ordering. This implies that the magnetic field ordering remains uniform across the emitting particle population.

\section{Discussion \& Conclusions}\label{sec:Discussion}

We have shown that the energy spectra and polarization degree from optical to X-rays for Mrk 421 can be modeled using a broken power-law electron energy distribution, where all the electrons experience the same magnetic field configuration. This is in contrast to Mrk 501, where the high-energy X-ray producing electrons were required to be in the presence of a more ordered field than the low-energy optical producing ones. The difference here is that the break energy in the energy spectra for Mrk 421 lies below the IXPE energy range, while for Mrk 501, it was at a higher energy. For Mrk 421, the steeper spectrum after the break naturally allows for a higher polarization degree in X-rays compared to optical wavelengths. Moreover, for Mrk 421, the model fitting rules out any significant variation in the magnetic field ordering with electron energies and is particularly inconsistent with the variation inferred for Mrk 501 \citep{Misra_2025}. This indicates that different sources may have different magnetic field stratification; indeed, it is possible that the same source may exhibit different behavior.

In this work, it is also shown that other particle energy distributions, such as the log-parabola and EDD, can also explain both the energy spectrum and the polarization degree in the optical and X-ray bands without invoking any dependence between the electron energy and magnetic field ordering. This is consistent with the results obtained for Mrk 501. Thus, if the particle distribution is more complex than a BPL, the same scenario can explain both Mrk 501 and Mrk 421. However, as discussed in \cite{Misra_2025}, evidence that this is indeed the case would require multiple spectro-polarimetric observations of a source to check if the polarization degree variation with spectral parameters is as predicted by these models.

One important thing to note is that if the emissions from different wavebands arise from the same region, then, to the first order, one should expect their polarization angles to be the same. For Mrk 421, the polarization angles for the different wavebands are, optical: $200 \pm 1^\circ$, and X-ray: $220 \pm 4^\circ$, where the values and errors are taken from \cite{2023NatAs...7.1245D}. It is not clear whether the differences in the polarization angles are significant and whether rapid variability in the polarization angle may lead to such differences.

During a latter June 4-9 observation of Mrk 421 the X-ray polarization angle varied systematically implying that the ordered magnetic field is probably of a helical nature and the variation occurs as the emission region moves along the field line \citep{2023NatAs...7.1245D}. As discussed in \cite{Misra_2025}, it will be interesting to study a source where the PD variation can be shown to be correlated with the spectral parameters, as expected for synchrotron emission. However, PD could also vary due to other factors, such as change in the magnitude of the ordered magnetic field.

These results indicate the need for more multi-wavelength spectro-polarimetric studies of HBLs, which will provide deeper insights into particle acceleration mechanisms and magnetic field dynamics in relativistic jets.

\section*{Acknowledgements}
We acknowledge the use of public data from the \textit{NuSTAR}, \textit{Swift}-XRT/UVOT from NASA’s High Energy Astrophysics Science Archive Research Center (HEASARC). R.M. thanks Prof. Shiv Sethi for useful discussions on the expected polarization from Synchrotron emission. H.B. and R.G. would like to acknowledge IUCAA for their support and hospitality through their associateship program.

\bibliography{sample701}{}
\bibliographystyle{aasjournal}



\end{document}